# Effects of skull properties on continuous-wave transcranial focused ultrasound transmission

Han Li,[1,3] Isla Barnard,[1] Tyler Halliwell,[1] Xinyu Zhang,[2] Andreas Melzer,[2,4] and Zhihong Huang[3]

[1] School of Science and Engineering, University of Dundee, Dundee, DD1 4HN, UK

[2] School of Medicine, University of Dundee, Dundee, DD1 9SY, UK

[3] School of Physics, Engineering and Technology, University of York, York, YO10 5DD, UK

[4] Institute for Computer Assisted Surgery, University Leipzig, Semmelweisstraße 14, 04103 Leipzig, Germany

ABSTRACT: Transcranial low-intensity focused ultrasound can deliver energy to the brain in a minimally invasive manner for neuromodulation applications. However, continuous sonication through the skull introduces significant wave interactions, complicating precise energy delivery to the target.

We present a comprehensive examination of intracranial acoustic fields generated by focused ultrasound transducers and assess the characteristics of cranial bone that affect acoustic transmission. Acoustic field maps were generated at 88 regions of interest across 10 historical and 2 Thiel-embalmed human skull specimens with sonication at frequencies of 220 kHz, 650 kHz, and 1000 kHz. The average peak pressure insertion losses for historical were $3.6\pm3.4$ dB, $9.3\pm3.3$ dB, and $14.8\pm5.8$ dB, respectively, and for Thiel skulls, the respective losses were $2.9\pm1.8$ dB, $9.4\pm2.6$ dB, and $17.0\pm5.5$ dB. The effect of skull thickness, skull density ratio, and skull curvature on intracranial peak pressure, power and focal area was investigated and linear fits produced. Several unfavorable focusing performances were observed in regions with excessive thickness variation. The effects of angulation and spacing between the transducer and the skull were also investigated. Preliminary findings indicate that wave superposition resulting from skull and transducer spacing could lead to a 30-40% uncertainty in peak recorded intracranial pressure.

# I.    INTRODUCTION

Transcranial Low Intensity Focused Ultrasound (LIFU) for neuromodulation is a promising and rapidly developing technology[1–6] with the ability to probe motor and cognitive functional connectivity in the brain in a non-invasive reversible manner.[7–11]

Transcranial LIFU offers several advantages over established non-invasive forms of neuromodulation. LIFU affords the possibility of neuromodulating brain tissue with the spatial precision required to translate to clinically meaningful targets.[2,5] LIFU also has the potential to neuromodulate deep brain structures,[3,6] whereas established modalities, such as transcranial magnetic stimulation (TMS) are primarily limited to cortical regions and have lower spatial precision.[12]

Though the mechanisms underlying this have been a source of debate, it is recognised that LIFU temporarily disrupts brain circuit function by altering the state of ion channels in neurons.[13] Recent in-vitro studies indicate that calcium-permeable ion channels require a continuous excitation for 200 ms and a constant 15 W/cm$^2$ energy deposition to activate a neuronal response.[13]Although in-vivo parameters have not yet been established, research suggests that a minimum pressure amplitude at specific frequency is typically required to elicit neural activity, and exceeding certain thresholds may induce adverse effects.[14] Variability in neuron types and regional tissue stiffness also likely influences responsiveness to different pressure levels.[15,16] Therefore, precise control of on-site pressure level is essential for assessing biophysical effects.

However, analogous to difficulties surrounding utilising ultrasound for brain imaging, it is a challenge to deliver LIFU through the skull, into the brain.[2] The skull imposes significant but not precisely defined attenuation on the acoustic field, making it challenging to deliver the desired pressure level and maintain target engagement (ensuring the intended target is stimulated while avoiding off-target unintended neuromodulation). Some in-vivo studies assume a uniform insertion loss across individuals or apply free-field pressure measurements as a substitute for in-vivo parameters,[17,18] despite the fact that this reduction can vary across subjects due to differences in skull properties and frequencies.[19] Such variability may compromise the accuracy of determining optimal ultrasound neuromodulation parameters.

Previous studies have extensively explored and characterized the effects of human skulls on acoustic transmission using pulses of a few cycles, revealing a strong inverse relationship between the acoustic pressure attenuation and skull thickness, as well as ultrasound frequency.[20–25] During transcranial LIFU treatment, pulse patterns contain several hundred cycles per pulse delivered, and as such would be considered to fall in the 'continuous-wave' regime in acoustics.

Such waveforms could introduce constructive or destructive interference at both the transducer-skull spacing and within the skull thickness. The combined effect may result in insertion losses that differ from the attenuation measurements obtained via a short-pulsed waveform.[26–28] Therefore, using skull thickness alone is insufficient to predict total energy loss during penetration. In Magnetic Resonance Imaging-guided Focused Ultrasound (MRgFUS) treatment for essential tremor,[29] the skull density ratio (SDR), which reflects contrasts between skull layers, influences the maximal energy required to achieve a therapeutic lesion and showing a correlation with treatment efficacy.[30] This implies that pressure levels are affected by the SDR value.

Ultrasound simulation software plays a crucial role in predicting the intracranial acoustic field,[31,32] studies using the k-Wave toolbox for transcranial FUS simulation highlight the importance of accurately modelling skull morphology, as the skull's shape significantly influences the resulting acoustic field.

In this study, we aim to emulate the conditions under which transcranial LIFU is delivered in practice, using continuous waveform at sub-MHz frequencies. Our intention is to assess the significance of the physical properties of the skull on focused ultrasound transmission, specifically the thickness, SDR and morphology, as well as skull-transducer placement. By comparing the acoustic field measurements in the free-field to that in the presence of a skull, we seek to identify the skull characteristics that may impede effective transmission and we aim to reveal the correlations between skull properties and insertion loss in terms of pressure and power, acoustic beam focusing performance, and targeting accuracy.

## II. MATERIALS AND METHODS

### A. Skull selection and physical properties

In this experiment, 10 historical human skulls (8 calvaria and 2 half-skulls) and 2 Thiel-embalmed[33] human calvaria were sourced from the Centre for Anatomy and Human Identification (CAHID). Both skull types are commonly accepted for experimental use, with minor discrepancies in bulk acoustic speed.[34] The historical specimens were visually inspected and selected based on variation in thickness, density, size and shape from a pool of approximately 200 available skulls. Although the provenance of the collection is not known, the extensive range and variability suggest a broad representation of ages and sexes.

The Thiel calvaria samples provided came from Thiel soft embalmed cadavers (most of the donors are between the ages of 75 and 95). These samples retain the elasticity and feel of living tissue, with both scalp and dura intact. Although the embalming process may induce changes in electrical conductivity, viscoelastic relaxation behaviour, and stiffness of the tissue, these changes are unlikely to significantly alter the ultrasound transmission characteristics.[35] The specimens were handled confidentially and the procedures involving the Thiel calvaria strictly followed the Anatomy Act 1984 and the Human Tissue (Scotland) Act 2006.[36]

For each half skull, 4 regions of interest (ROIs) were selected to correspond with specific brain targets relevant to clinical applications.[37–40] For the calvaria samples (both historical and Thiel) these regions were mirrored across the medial line, giving 8 ROIs for each calvaria. This resulted in a total of 88 ROIs (72 in historical dry skulls and 16 in Thiel). FIG. 1 (a); details how ROIs are distributed across a half skull sample. To obtain structural and morphological information from the skulls, a clinical CT scanner (GE Revolution EVO CT scanner; with in-plane spacing at 0.44 mm for both axes and slice thickness of 0.625 mm with 0.31 mm interval) was used, this allows that findings are directly translatable to clinical applications. Prior to CT scanning, ROIs were marked on each sample using metallic ball bearings to identify ROIs during image analysis. Two examples of the resulting ROI volume obtained from a CT scan are shown in FIG. 1 ($b_1$ and $c_1$). The volumes are used as illustrative examples throughout and are labeled as S1 and S2.

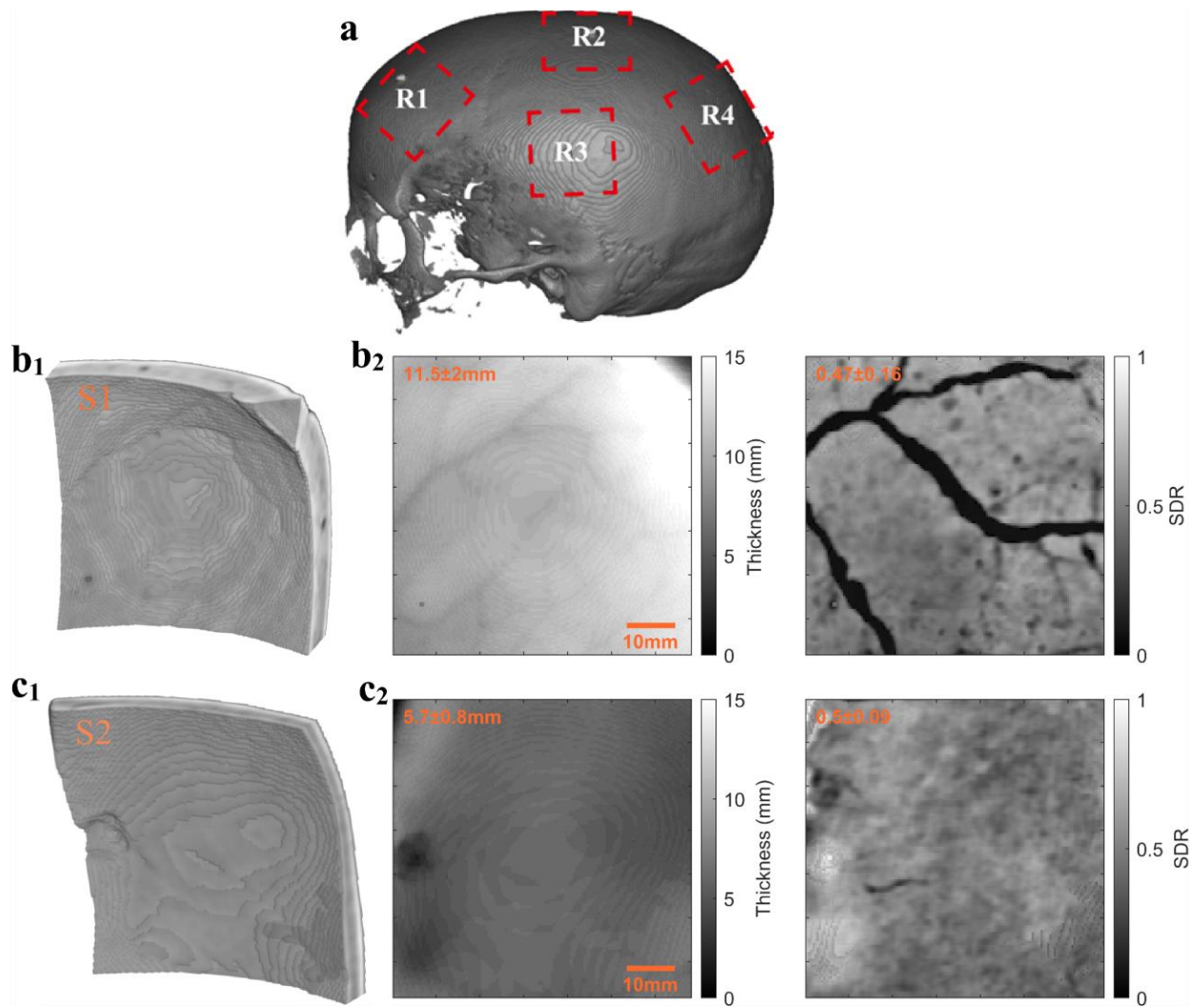

FIG. 1. (Color online). A representative skull with ROIs details. Two representative regions were chosen for the following analysis. (a) ROI locations on a half skull; R represent region of interests that are applicable across all skulls. R1 is frontal bone, R2, R3, and R4 are superior parietal, lateral parietal (partial temporal in some samples), and posterior parietal bone, respectively. ($b_1$) The CT volume of an example ROI (60 mm by 60 mm surface area) labeled throughout as S1; S represent specific sample regions selected from region of interests for illustrative purposes. ($b_2$) Thickness and SDR map of S1. Average thickness and SDR are 11.5±2.0 mm, and 0.47±0.16, respectively. The radius of curvature (RoC) of S1 is approximately 55.1 mm. ($c_1$, $c_2$) Example of a ROI labeled as S2. Average thickness, SDR and RoC of are 5.7±0.8 mm, 0.5±0.09, and 81.6 mm, respectively.

ROIs were characterized by thickness, skull density ratio (SDR) and radius of curvature (RoC), all obtained from analysis of the CT scans. The SDR is used as a screening and planning tool in MRI-guided focused ultrasound neurosurgery and is the ratio between the minimum and maximum radiodensity;[41] which for skulls is a metric representing the ratio between trabecular and cortical bone.

To characterize the skulls, each CT scan was imported into MATLAB. A surface casting method (an extension to the ray-casting method[42,43] used to find SDR from a CT scan); that considers ultrasound insertion angle, was developed to calculate both the SDR and the thickness.

Each volume was defined by selecting an approximately 60 mm by 60 mm square region from the skull surface, with the ROI metallic marker at the center. Each segmented ROI was 136 voxels wide and 193 voxels long. The ROI marker was removed from the image; and the inner and outer surfaces were defined as the point where HU units reach 300 HU.

A straight line was drawn from each outer surface voxel to the inner surface. The length of each line was multiplied by a cosine factor of the predicted sound insertion angle, defined as the ratio of the vertical height of the outer surface pixel relative to the transducer geometrical focus to the acoustic path length. In practice, the geometrical focus was located 65 mm deep from centre of the ROI. These adjusted thickness values were then averaged to calculate the mean and standard deviation of the thickness for the ROI. This adjustment ensures that the effective thickness reflects the actual distance travelled by the acoustic wave, eliminating biases introduced by the viewing orientation. However, the approach is transducer-specific and may result in variation in the effective thickness when using different transducers or configurations. The SDR was calculated based on the same principle (without angulation correction). To recover a realistic density for trabecular bone, a lower bound threshold was set to +100 HU to compensate for density reductions caused by re-entered air during the CT-scan. FIG. 1 (b$_2$ and c$_2$) show the corresponding thickness and SDR maps of S1 and S2.

To determine the curvature of the ROI and account for morphology differences between both sides, the inner and outer surface of the ROI were fitted to a best-fit spherical surface. The average of the two curvatures was used as the metric to indicate RoC.

The averaged thickness, SDR, and RoC for historical ROIs were 6.9±2.1 mm, 0.41±0.21, and 63.3±21.6 mm, respectively, and for Thiel ROIs, the respective values are 7.9±2.4 mm, 0.55±0.14, and 59.6±28.1 mm. The three characteristics of the grouped four regions data are presented in TABLE I.

TABLE I. The mean and standard deviation of skull region-specific parameters over all 88 ROIs tested.

|                 | R1        | R2        | R3        | R4        |
| --------------- | --------- | --------- | --------- | --------- |
| Thickness (mm)  | 7.7±2.6   | 6.4±1.9   | 5.8±1.9   | 8.0±2.3   |
| SDR             | 0.42±0.23 | 0.46±0.19 | 0.47±0.21 | 0.40±0.19 |
| RoC (mm) | 54.6±10.5 | 80.3±17.5 | 48.5±12.6 | 51.2±7.7 |

### B. Experimental setup & acoustic measurements

Three single element focused ultrasound transducers with centre frequencies of 220 kHz, 650 kHz, and 1000 kHz, with identical diameters of 60 mm and focal lengths of 75 mm, (Precision acoustics, UK) were selected. Transducers were driven by approximately 0.5 ms sinusoidal waveform generated by a signal generator (33500B, Keysight, US), which produces peak negative pressure in the free-field of 0.06 MPa at 220 kHz, 0.25 MPa at 650 kHz, and of 0.35 MPa at 1000 kHz. A 1 mm needle hydrophone (Precision Acoustics, UK) was used to capture the acoustic signal. Prior to the experiment, hydrophone sensitivity and power measurement calibrations were performed in accordance with hydrophone characterization standards IEC 62127-1/2,[44,45] the transducer transmitted field measurement standard IEC 61828,[46] and the ultrasonic power measurement standards IEC 61689[47] and IEC 61161.[48]

The hydrophone was mounted on a three-dimensional motor stage (Velmex, US), and was positioned at the focal depth, pointing towards the transducer. The received signal was sampled at 40 MHz by a PC oscilloscope (Picoscope, UK), and the waveform was reconstructed through a center frequency band-pass filter with a gate size of 10 Hz. The amplitude of the signal was then extracted by applying Fast Fourier Transformation to 6 to 25 cycles of steady-state waveform at 220 kHz to 1000 kHz. Free-field mapping with a step resolution of 0.5 mm was performed as a reference for each frequency. The sizes of the transverse mapping planes were 30 mm by 30 mm (220 kHz) and 20 mm by 20 mm (for 650 kHz and 1000 kHz). The areas chosen sufficiently to cover the full width half maximum (FWHM) beam area, and importantly are small enough to afford daily calibration prior to experimental data acquisition.

Prior to the experimental setup as shown in FIG. 2, each historical skull sample was degassed in a vacuum tank until no visible air bubbles escaped from the incised edge. The Thiel embalmed skulls were also degassed as above. Following degassing, Thiel skulls were vacuum sealed using an acoustically & visually transparent sealing bag (with tested transmission efficiency > 99%). Sealing prevented Thiel embalming fluid leaking from the sample into the water tank, and also prevented detachment of soft tissues from the skull caused by water ingress.

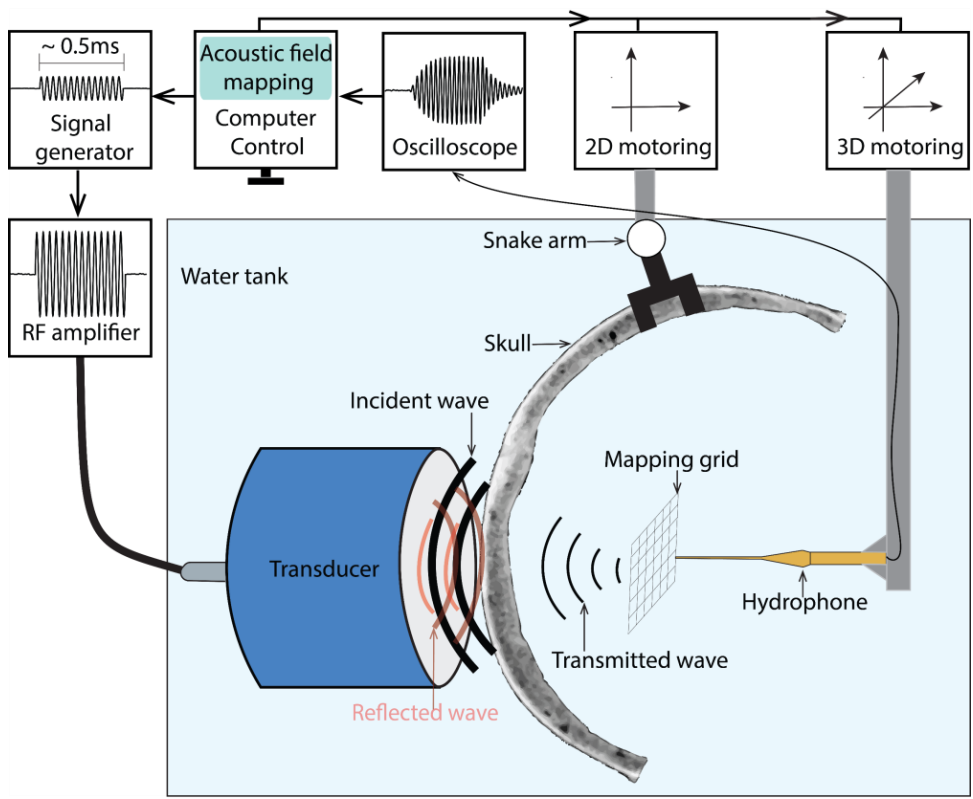

FIG. 2. (Color online). Experimental setup of the transcranial acoustic field measurement system.

The degassed skull was then clamped on a snake arm that can be locked in any position, and which was fixed on a two-dimensional motoring stage, and positioned between the transducer and hydrophone.

To investigate the influence of the skull properties on the transmitted pressure and power level (and the sharpness of the focus), the position of all ROIs under testing were initially aligned to near-normal incidence, maneuvered to approximately 10 mm from the transducer aperture, and then slightly adjusted to where constructive wave patterns were clearly observed. The hydrophone was located at the local maxima along the axial direction, and the acoustic field measurements were taken in the same lateral position as the free-field mapping to capture skull-induced changes in the transmitted acoustic field.

To investigate the impact of transducer-skull spacing on the transmitted acoustic pressure level, using a single ROI, the gap between the skull and the transducer aperture was increased from 10 mm to 50 mm, in increments of 0.2 mm, and the pressure at the focal point was recorded. Using the 650 kHz transducer and a single ROI, the effect of angle between the transducer and the skull was also investigated. The relative angle was moved from 0° to 15°, then to 30°, and the resulting field was mapped with a step resolution of 0.5 mm along the beam axis. The corresponding pressure $p$ at each sampling position and the total acoustic $P_{HP}$ power can be calculated by using equation:[46]

$$P_{HP} = \iint \left( \overline{p(x,y)^2} / \rho c \right) dy dx, \ p = U / M_f \tag{1}$$

Where $U$ is the signal amplitude, $M_f$ is the hydrophone sensitivity at transducer working frequency, $\overline{p(x,y)}$ is the time averaged acoustic pressure at coordinate $(x, y)$, $\rho$ and $c$ are the density and sound velocity of water (1000 kg/m³ and 1480 m/s, respectively), $dy dx$ is the pixel area (assumed to be the square of the step-resolution). By integrating over the mapping area, we recover the $P_{HP}$; the approximate total acoustic power transmitted through that area.

## C. Acoustic field analysis

The scheme used to analyze the acoustic field is shown in FIG. 3, using volume S1 and S2 (FIG. 1) as examples. In order to compare the transmitted acoustic field across 88 ROIs, the intracranial focal area was defined as the region where the pressure level above the -6 dB (half-maximum) and is shown in FIG. 3 ($a_1$, $b_1$ and $c_1$) as the dashed circle on the maps.

The aberration of the beam was further assessed by quantifying the normalized pressure-weighted phase errors ($pe$), which combine information from the pressure and phase maps,[49] by comparing the intracranial focal area and the corresponding area projected onto the reference, using equations:

$$p_{free,inst}(x, y, \psi_{free}) = p_{free}(x, y) \cos(\psi_{free}(x,y)) \tag{2}$$

$$p_{IC,inst}(x, y, \psi_{IC}) = p_{IC}(x, y) \cos(\psi_{IC}(x,y)) \tag{3}$$

$$\Delta p(x, y) = \frac{1}{N_\psi} \sum_{N_\psi} |p_{free,inst}(x, y, \psi_{free}) - p_{IC,inst}(x, y, \psi_{IC})|, \qquad \psi \in [0, 2\pi] \tag{4}$$

$$pe = \frac{1}{N_{IC,focal}} \left( \sum_{N_{IC,focal}} \Delta p(x, y) \right) * 100\%, \qquad (x, y) \subset IC, focal \tag{5}$$

Where $p_{free,inst}(x, y, \psi_{free})$ and $p_{IC,inst}(x, y, \psi_{IC})$ represent the instantaneous pressures at coordinate $(x, y)$ and phase angle $\psi_{free}$ and $\psi_{IC}$ in the free-field and intracranial space, respectively. $p_{free}(x, y)$ and $p_{IC}(x, y)$ denote the pressure amplitudes at $(x, y)$, normalized to the peak values in their respective field. $N_\psi$ indicates the number of phase samples, spanning from 0° to $2\pi$ in increments of 1°. $\Delta p(x, y)$ is the average pressure-weighted phase difference at $(x, y)$, and $N_{IC,focal}$ refers to the number of sampled points in the intracranial focal area. The phase error $pe$, quantifies the average difference of the two fields, specifically focusing to the intracranial focal area. The two fields were synchronized at the moment their peak pressure was reached.

As depicted in FIG. 3 ($a_2$-$c_2$ and $a_3$-$c_3$), the colorless grids represent the peak (0°, corresponding to the point where peak pressure occurred) and ambient pressure ($\pi/2$) fields in the free-field, and the acoustic fields transmitted through S1 and S2 were represented by the color grid. The full-cycle pressure vibration error compared to the free-field at each frequency is shown in FIG. 3 ($a_4$-$c_4$). This method allows for the comparison of the two wave propagation patterns at the focal zone and visualizes the extent of the wave distortion, providing deeper insights than a simple pressure field comparison.

The system control, data acquisition and data processing software were developed in-house in MATLAB, and the acoustic field aberration analysis was performed in SPSS V26.0.

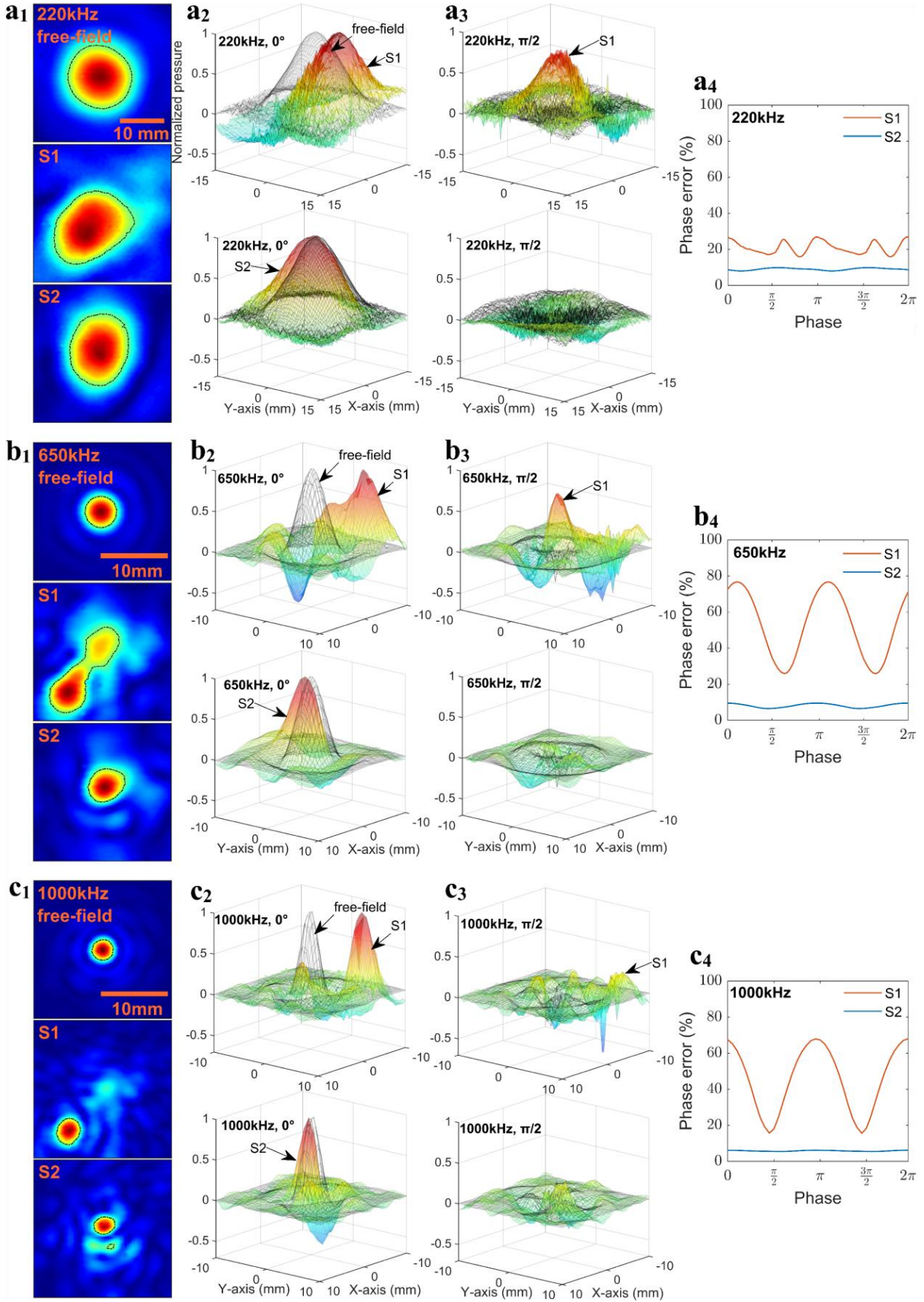

FIG. 3. (Color online). Acoustic field aberration analysis for the representative regions S1 and S2 that was described in FIG. 1 ($a_1$-$c_1$) Cross-sectional beam area pressure maps at the focal depths of the 220 kHz, 650 kHz and 1000 kHz transducers, produced at the focal depth in both free-field and with the presence of S1 and S2. ($a_2$-$c_2$, $a_3$-$c_3$) Color-less grids of three frequencies show a normalized pressure map at the focal depth in free-field at phase $=0°$ and $\pi/2$, respectively. The composed color images are those with the presence of S1 and S2. ($a_4$-$c_4$) The full vibration cycle phase errors of the pressure fields of S1 and S2 compared to the reference free-field were recorded at three frequencies.

## III. RESULTS

### A. Transcranial pressure and power level, and sharpness of the focus analysis

The average results from historical and Thiel ROIs are presented in **Error! Reference source not found.**. The results from historical ROIs were aggregated, and fitted to a linear regression model, accompanied by a 95% confidence interval (CI), to demonstrate the correlation between the pressure and power level, as well as the sharpness of the focus, to the ROI thickness (FIG. 4), SDR (FIG. 5) and the RoC (FIG. 6) at the three frequencies. To evaluate the effect of skull type, Thiel ROIs (Thiel) data were overlaid onto the plots for direct comparison.

TABLE II. The average results of the three frequencies from historical and Thiel ROIs.

| Historical/ Thiel | 220 kHz | 650 kHz | 1000 kHz |
|---|---|---|---|
| Normalized peak pressure (dB) | -3.6±3.4 | -9.3±3.3 | -14.8±5.8 |
| | -2.9±1.8 | -9.4±2.6 | -17.0±5.5 |
| Normalized power (dB) | -5.8±5.7 | -15.2±6.0 | -24.2±9.0 |
|  | -3.6±1.8 | -14.6±2.9 | -26.1±8.9 |
| Normalized focal area | 1.3±0.3 | 1.4±0.3 | 1.4±0.4 |
|  | 1.4±0.7 | 1.8±0.8 | 2.7±1.9 |
| Phase error (%) | 15.5±4.0 | 16.3±4.8 | 21.8±10.5 |
|  | 13.7±7.0 | 20.7±9.7 | 28.4±11.4 |
| Focus shift (mm) | 2.3±1.9 | 1.1±1.1 | 0.9±1.2 |
|  | 2.0±1.5 | 0.8±1.1 | 1.3±1.0 |

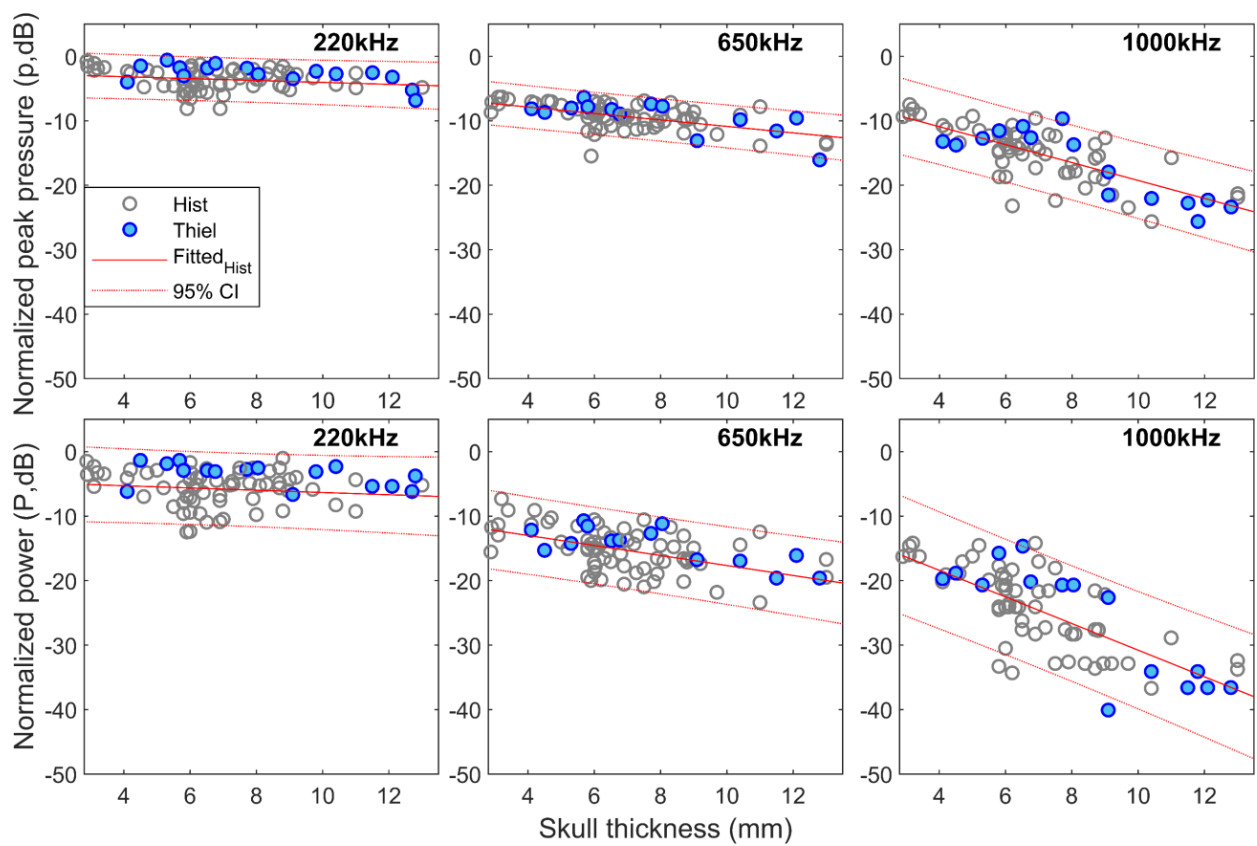

FIG. 4. (Color online). Skull thickness versus the normalized transcranial peak pressure level (dB), power level (dB) at three frequencies. The grey circles indicate historical skull ROIs (Hist), and filled blue circles represent Thiel ROIs (Thiel). The linear fit (plus 95 % confidence interval, CI) to the historical ROI data (Fitted_Hist) is indicated as shown.

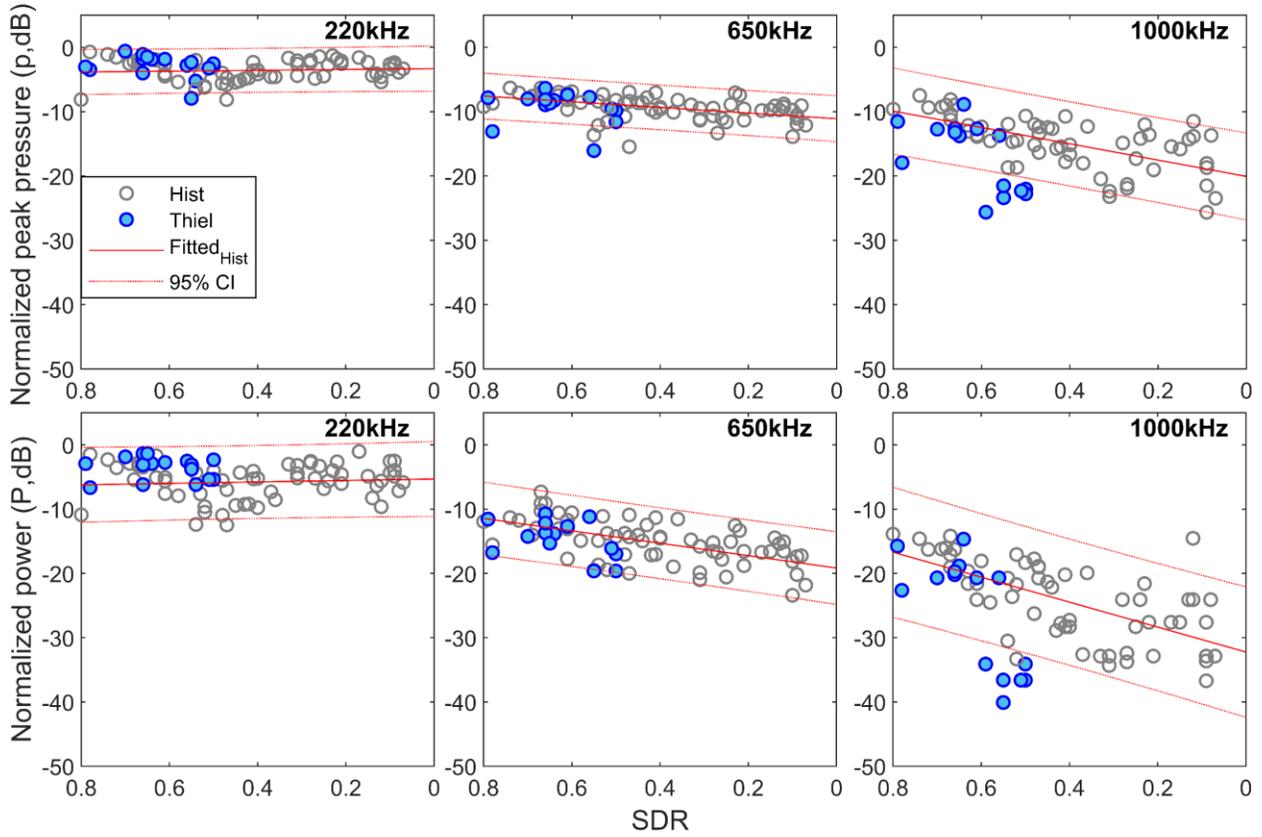

FIG. 5. (Color online). Skull SDR versus the normalized transcranial peak pressure level (dB) and power level (dB).

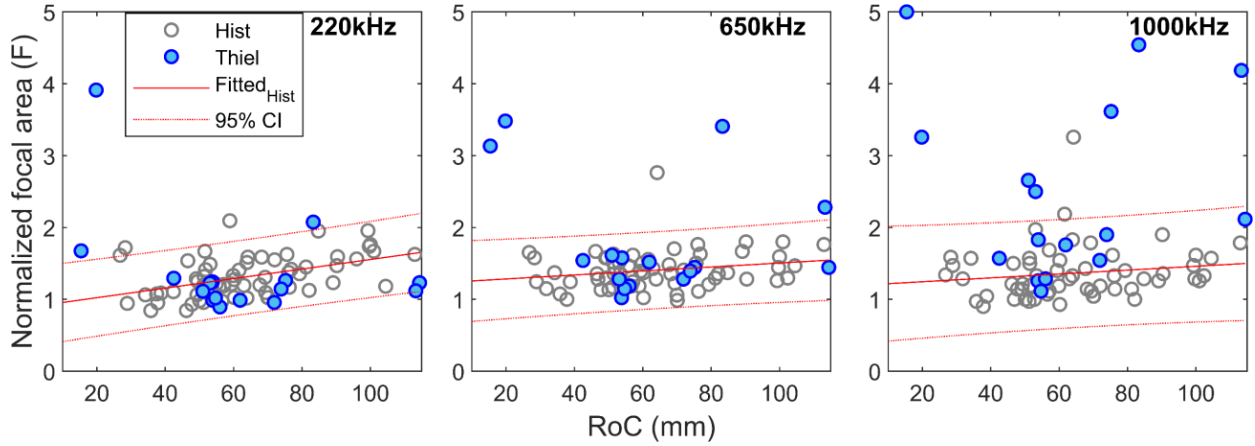

FIG. 6. (Color online). Skull ROI RoC versus the normalized focal area (F) at three frequencies. The focal areas were normalized to the free-field reference- a larger ratio indicates a blurred focus, and a smaller ratio indicates a sharper focus.

FIG. 4 illustrates the relationship between insertion loss levels of pressure and power and average ROI thickness. At 220 kHz, insertion loss for historical ROIs ranged from -0.6 dB to -8.1 dB, and -1.0 dB to -12.5 dB for peak pressure and power, respectively, relative to free-field values. Thiel ROIs exhibited smaller ranges of -0.6 dB to -6.8 dB, and -1.4 dB to -6.7 dB, respectively. These values appeared to be independent of thickness variations.

At 650 kHz, historical ROI insertion loss increased to range from -6.4 dB to -15.5 dB for peak pressure and -7.3 dB to -23.4 dB for power. For Thiel ROIs, the corresponding ranges were -6.4 dB to -16.1 dB, and -10.7 dB and -19.6 dB. The correlation between these insertion loss levels and thickness (TH, mm) were discernible, and fitting revealed the linear regression equations:

$$p_{Hist,650kHz} = -0.5TH - 5.9 \pm 3.3, (R^2 = 0.30)$$

$$P_{Hist,650kHz} = -0.8TH - 9.9 \pm 6.0, (R^2 = 0.24)$$

$$p_{Thiel,650kHz} = -0.6TH - 4.2 \pm 4.3, (R^2 = 0.52)$$

$$P_{Thiel,650kHz} = -0.8TH - 8.6 \pm 4.7, (R^2 = 0.57)$$

Where $p$ and $P$ are pressure and power level, respectively. At the higher frequency of 1000 kHz, insertion loss further increased. Historical ROI values ranged from -7.5 dB to -25.6 dB, and -14.2 dB to -36.7 dB, respectively, Thiel ROIs exhibited similar ranges of -8.9 dB to -25.6 dB, and -14.7 dB to -40.1 dB. The correlations with thickness were more pronounced, as expressed by the equations:

$$p_{Hist,1000kHz} = -1.4TH - 5.4 \pm 5.8, (R^2 = 0.52)$$

$$P_{Hist,1000kHz} = -2.1TH - 10.2 \pm 9.0, (R^2 = 0.50)$$

$$p_{Thiel,1000kHz} = -1.7TH - 3.1 \pm 6.2, (R^2 = 0.77)$$

$$P_{Thiel,1000kHz} = -2.6TH - 4.2 \pm 11.4, (R^2 = 0.69)$$

Notably, the data collected from the Thiel ROIs mostly fell within the confidence intervals for historical ROIs at all three frequencies, highlighting comparable trends.

Similar patterns were also observed concerning the correlations between SDR and the insertion loss. As shown in FIG. 5, at 220 kHz, the insertion loss of pressure and power exhibiting independence from variations in SDR, whereas weak correlations were noticed at both 650 kHz and 1000 kHz for historical ROIs:

$$p_{Hist,650kHz} = 4.4SDR - 11.1 \pm 3.5, (R^2 = 0.22)$$

$$P_{Hist,650kHz} = 9.6SDR - 19.1 \pm 5.5, (R^2 = 0.35)$$

$$p_{Hist,1000kHz} = 12.7SDR - 20.1 \pm 6.6, (R^2 = 0.39)$$

$$P_{Hist,1000kHz} = 19.4SDR - 32.2 \pm 9.9, (R^2 = 0.40)$$

Thiel ROI data mostly aligned with the confidence intervals at 220 kHz, whereas several deviations were apparent at 650 kHz and 1000 kHz. The limited SDR range in Thiel ROIs restricted comprehensive analysis of its correlation with insertion loss.

In investigating the effect of skull curvature on bending ultrasound wave transmission, and its influences on the sharpness of the focus for the specific transducer used, FIG. 6 shows that at 220 kHz, the focal area for historical ROIs varying between a factor of 0.8 to 2.1 relative to the free-field, demonstrating a weak correlation with the skull curvature (RoC, mm):

$$F_{Hist,220kHz} = 6.6 * 10^{-3} RoC + 0.9 \pm 0.5, (R^2 = 0.22)$$

Where $F$ is the intracranial focal area normalized to free-field. However, the focal area varied by a factor of 0.9 to 2.7 for 650 kHz and by 0.9 to 3.3 for 1000 kHz, with no discernible correlation to the skull curvature. Although most historical data align with confidence interval, deviations were more pronounced in Thiel ROIs, particularly at higher frequency.

### B. ROIs associated with poor ultrasound transmission performance

As shown in FIG. 6, some acoustic fields exhibited significantly expanded focal area, which may also accompany a strong focus shift. As two representative ROIs results that illustrated in FIG. 3 ($a_1$-$c_1$), after transmitted through S1, compared to the free-field, the intracranial foci shifted 5.8 mm from the center of the focal depth at 220 kHz, and 7.8 mm at both 650 kHz and 1000 kHz. The focal area expanded by a factor of 1.1, 3.3, and 1.3 at 220 kHz, 650 kHz, and 1000 kHz, respectively. At 220 kHz, phase errors between the two overlapping grids within S1-focal area were 23.3% at peak pressure and 21.7% at ambient pressure. For 650 kHz and 1000 kHz, larger focal shifts resulted in significant deviation in pressure and phase in the S1-focal area with errors of 76.8% and 68.0% at peak pressure, and 26.0% and 15.6% at ambient pressure, respectively. The overall phase errors across a full vibration cycle were 20.9% at 220 kHz, 53.4% at 650 kHz and 46.4% at 1000 kHz.

In comparison, the acoustic field after being transmitted through S2 region exhibited less aberration. The focal shifts were limited to 0.7 mm at 220 kHz and 650 kHz, and 0.5 mm at 1000 kHz. The focal area deviated by less than 5% from the free-field value, with average phase error of 9.0%, 8.1% and 5.5% at three frequencies, respectively.

These representative results suggest that skulls with characteristics similar to the S2 region are ideal for transcranial sonication, allowing straightforward planning without concerns about off-target sonication or inadequate acoustic focusing.  In contrast, acoustic field transmitted through ROIs such as S1 revealed severe aberrations, including reduced wave pattern coherence, significant focus shifts, and the absence of meaningful focal point.

To identify bone characteristics beyond average thickness and SDR that contribute to these undesired aberrations, all the phase maps were manually inspected, and phase error was fitted to a logistic regression model to define the receiver operating characteristic curve. The determined cut off phase errors were 20.2%, 21.5% and 30.6% for 220 kHz, 650 kHz and 1000 kHz, respectively. Analysis of ROIs exceeding these thresholds revealed a significant factor: excessive variation in thickness across the ROI. Specifically, the ROIs with phase error exceeding the cut-off exhibited thickness deviations of at least 2.0 mm, 1.6 mm and 1.3 mm at 220 kHz, 650 kHz and 1000 kHz, respectively.

### C.  Transducer-skull spacing and placement angulation

During early testing of longer period sonication (dozens of cycles), a wave superposition effect started to be noticeable across the ROIs- an investigation is presented in FIG. 7.

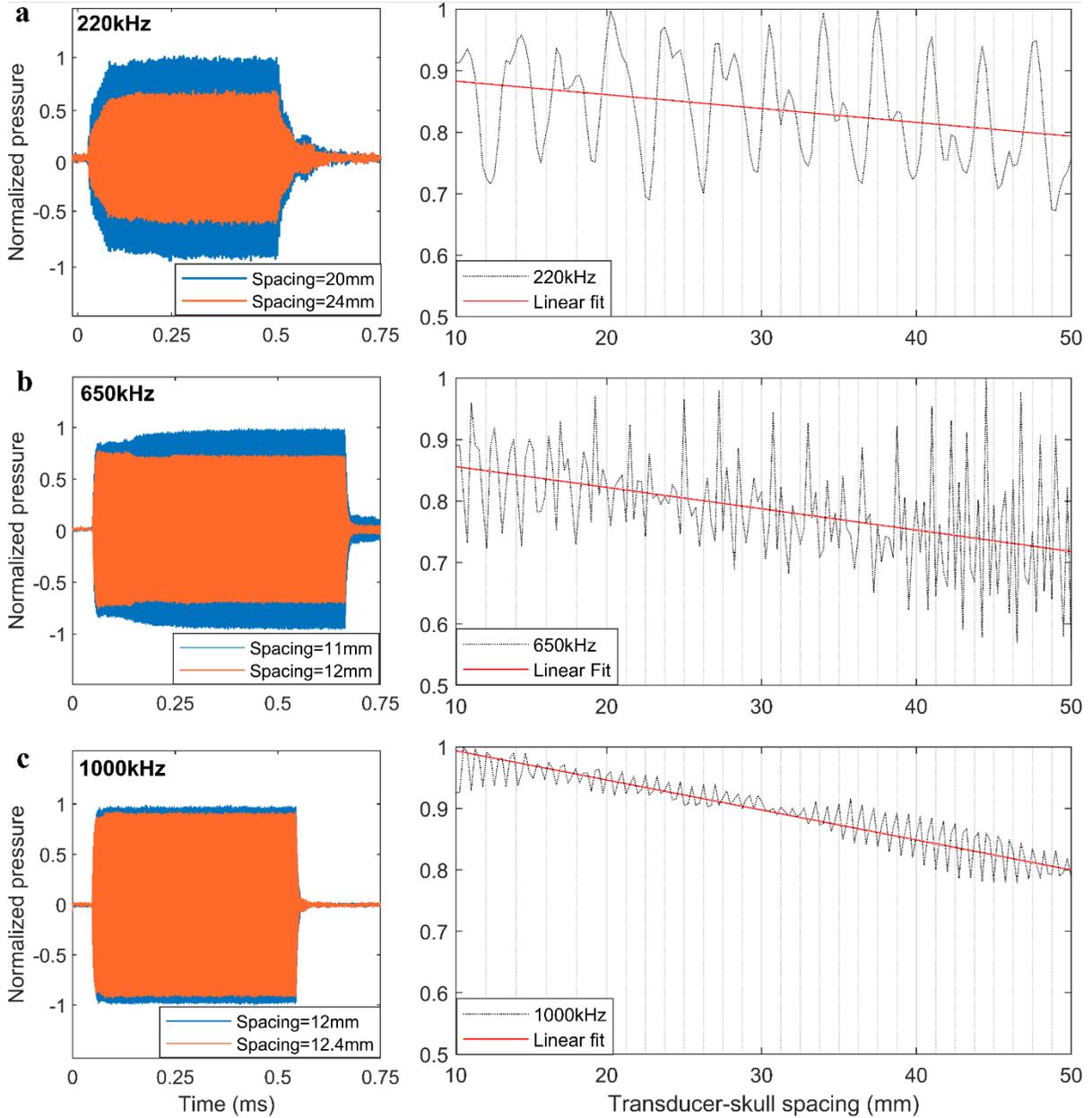

FIG. 7. (Color online). Pressure variation at the intracranial focus along with the increase of the transducer and representative skull (thickness of 3.2±0.8 mm, SDR of 0.65±0.25 and RoC of 70.0 mm) spacing at three frequencies and its waveform. (a-c) Waveform amplitude development at spacing approximately of 20 mm and 24 mm, 11 mm and 12 mm, 12 mm and 12.4 mm, and the pressure variation from 10 mm to 50 mm spacing at 220 kHz,650 kHz, and 1000 kHz, respectively. The linear fitted red line showed the overall decreasing trend of the pressure along the spacing.

The effects of a skull piece (thickness of 6.2±0.8 mm, SDR of 0.54±0.18, and RoC of 56.1 mm) on the acoustic field at normal incidence at 650 kHz, and the impact of its placement angle and resulting field aberration was tested and is illustrated in FIG. 8.

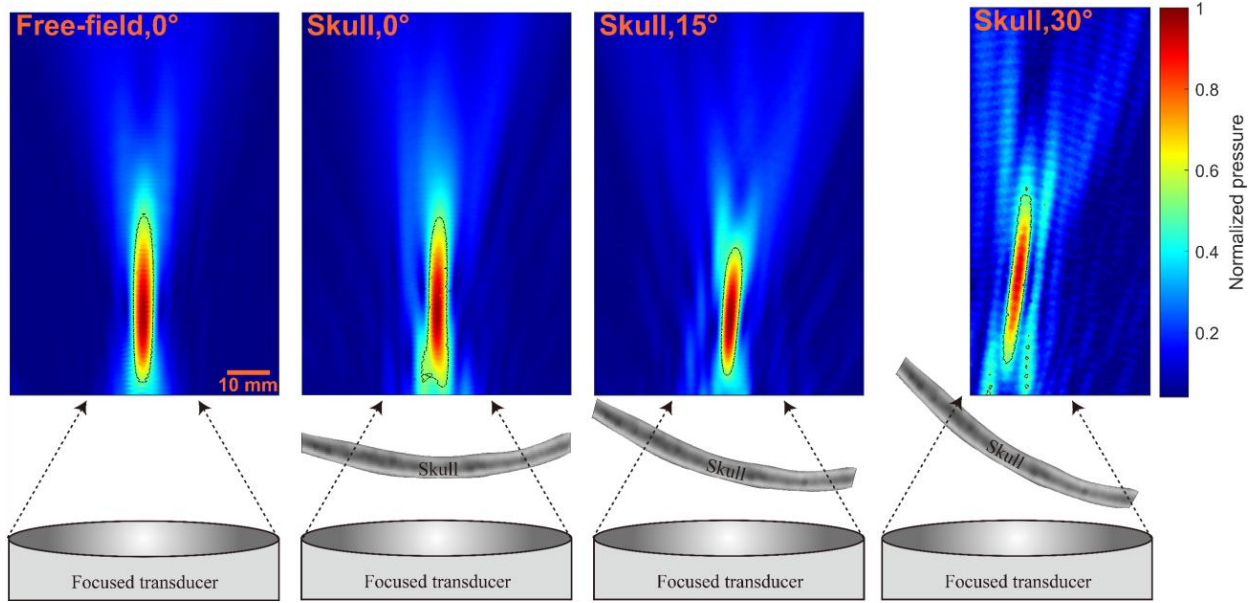

FIG. 8. (Color online). Acoustic field visualization of the 650kHz transducer beam profile in free-field and with the presence of a representative skull (thickness of 6.2±0.8 mm, SDR of 0.54±0.18, and RoC of 56.1 mm) at approximate angulation of 0°, 15° and 30° to the right side of the transducer aperture. Pressure is normalized to its maximum value, the area enclosed by the dashed circle is where the pressure is above -6 dB of the peak pressure. The maneuvering of the hydrophone was blocked by the cranial bone at a 30° alignment, resulting in a limited field of view.

## IV.    DISCUSSION

To accurately capture acoustic field variations during sonication, limits were set on the temporal and spatial scope of data acquisition.  The mapped transverse planes extended across at least twice the FWHM of the free-field focus. This provided adequate data collection in instances of focus shift or expansion, but also allowed field mapping to be completed within an hour. Axial scans were performed to visually inspect the severity of beam distortion caused by variations in transducer-skull angulation. Due to the time-intensive nature of this procedure, axial mapping was not applied to each ROIs. The pulse duration was set to approximately 0.5 ms. Though typical pulses delivered in in-person trials may extend beyond this range, [17,18] 0.5 ms is long enough for the characteristics of continuous-wave sonication to develop.

The selected historical ROIs spanned a comprehensive range of characteristics. Thicknesses ranged from 3.1 mm to 13.0 mm, SDR from 0.07 to 0.80 and RoC from 26.8 mm to 113.0 mm. The Thiel ROIs examined were limited to thicknesses of between 4.1 mm to 12.8 mm, SDR values between 0.48 and 0.79, and RoC values between 15.4 mm and 114.4 mm. By using different ROIs across the two Thiel samples, we were able to reveal similarities and differences between transmission across historical and embalmed specimens.

### A. tFUS transmission through historical and Thiel skull

The historical and Thiel results presented in **Error! Reference source not found.** show average transcranial peak pressure insertion loss of 3.6 dB and 2.9 dB at 220 kHz, 9.3 dB and 9.4 dB at 650 kHz, and 14.8 dB and 17.0 dB at 1000 kHz, respectively. In comparison, another study that tested frequencies of 150 kHz, 500 kHz, and 1000 kHz measured corresponding peak pressure attenuations of 6.6 dB, 8.9 dB, and 18.4 dB, respectively.[50] These attenuation values are extrapolated to be higher than those observed in our study.

The average transcranial power insertion loss for historical and Thiel ROIs demonstrated a less than square relationship to the pressure insertion loss, of 5.8 dB and 3.6 dB, 15.2 dB and 14.6 dB, and 24.2 dB and 26.1 dB, respectively. These findings may be evidence of spatial spreading of the acoustic energy, which also contributed to an increase in the focal area by a factor of 1.3 and 1.4, 1.4 and 1.8, and 1.4 and 2.7 at the corresponding frequencies. As illustrated in FIG. 3 ($a_1$-$c_1$), in a normalized scenario, the S1 area enclosed in the dashed line and its surrounding area both distributed more energy than the free-field- indicating that, in practice, a sonication may affect more brain tissue than intended in order to produce the same level of pressure at the brain target.

Overall, Thiel results aligned with historical data when correlated with thickness, with most data points fell within the confidence intervals. However. the historical SDR model demonstrated reduced predictive accuracy for insertion loss compared to thickness model, with more than 35% of Thiel results deviating from historical ROIs confidence interval at 1000 kHz.

The recorded aberration at different phases enables visualization of the coherence of the acoustic beam and enable quantitative comparison between two acoustic fields, although further investigation is required to fully understand the skull properties that causing these aberrations. The average transcranial phase errors for historical and Thiel ROIs were 15.5% and 13.7%, 16.3% and 20.7%, and 21.8% and 28.4% at the three frequencies, respectively. For context, the phase error in the free-field maps was in the range of 5%-8%, such variations might stem from measurement inconsistency due to finite sampling resolution and slight repositioning mismatch, however, empirically, errors below the cut-off values of 20.2%, 21.5% and 30.6% at the respective frequencies typically exhibited insignificant deviation from the free-field. Practically, the cut-off error tolerance increased with frequency, as even minor deviations could result in significant phase errors due to a more concentrated focal area and greater phase differences between adjacent points. This assertion is corroborated by the average focus shift results, which indicates the lateral displacement of peak pressure relative to the free-field focus: 2.3 mm and 2.0 mm, 1.1 mm and 0.8 mm, and 0.9 mm and 1.3 mm at the respective frequencies. Notably, the average shifts at 220 kHz were larger than those at 650 kHz and 1000 kHz, whereas exhibited lower average phase error. This is attributed to the shifts remaining below half of the wavelength and were maintained within the FWHM distance to the geometric focus.

Using a 10 mm thick historical skull as an example, our linear regression fit models as demonstrated above indicate minimum peak pressure insertion loss of 0.6 dB, 7.6 dB, and 13.6 dB, along with power insertion loss of 0.6 dB, 11.9 dB, and 22.2 dB at frequencies of 220 kHz, 650 kHz, and 1000 kHz, respectively. The insertion loss of the peak pressure at 1000 kHz falls below several studies provided values of the estimated threshold of 14.5 dB[28], 16.6 dB[51] and 18.4 dB[50] absorption loss of the skull. The insertion loss at 220 kHz is even below the reflection coefficient of 4.8 dB.[52] A conservative analytical study simulated skull attenuation under continuous wave conditions and reported minimum pressure reductions of 2.9 dB, 4.4 dB, and 5.5 dB at frequencies of 200 kHz, 600 kHz, and 1000 kHz, respectively.[53] In comparison, this study observed comparable reduction at 220kHz but higher reduction at 650 kHz and 1000 kHz. These findings align with the expectation that simulated pressure losses are lower than the experimental measurements.

Additionally, the average power losses for a 10 mm thick Thiel skull were approximated as 3.6 dB, 16.6 dB and 30.2 dB at 220 kHz, 650 kHz and 1000 kHz, respectively. In comparison, a study using freshly excised skulls measured average energy loss transmitted from the intracranial space to outward space with radiation force balance reported power attenuations of 2.9 dB, 20.1 dB, and 29.0 dB at frequencies of 270 kHz, 836 kHz and 1402 kHz, respectively,[19] showing minor discrepancies after extrapolation. These differences may arise from variations in experimental configurations and the condition of the skull specimens.

Consequently, if these linear fits be utilised to deliver continuous narrowband waveform for transcranial sonication, the upper bound of the confidence interval should be used for planning as a precautionary measure to mitigate the risk of excessive energy delivery to the target.

## B. Impact of skull morphology on tFUS transmission

Although thickness and SDR are considered primary factors influencing insertion losses, as shown in FIG. 6, acoustic transmission performance is also affected by skull morphology. The intracranial focal area exhibits an overall increasing trend along greater RoC, though its predictability decreases with increasing frequency. Quantitative analysis of phase error and ROI morphology characterises suggest that skull thickness variation plays a role in acoustic transmission. As illustrated in FIG. 1 and TABLE I, four regions, separated by the groove for superior sagittal sinus and coronal sutures were selected without pre-defined preferences.

The R1 region, situated within the frontal bone, are generally characterized by their thickness (7.7±2.6 mm), but it also features a smooth outer surface and comparable RoC (54.6±10.5 mm) to the transducer. In practice, waves propagating through this region exhibited minimal focus shift and phase aberration. However, two ROIs within the R1 region featuring a bulging inner table, with a curved outward inner surface, resulted in an erroneous calculated inward curvature (RoC of 15.4 mm and 19.8 mm, thickness deviation of 2.3 mm and 2.0 mm). The transmitted sound beam exhibited significant phase aberration at all three frequencies resulting in average focal area increased by a factor of 2.6, 3.3, and 4.1, with corresponding average phase errors of 25.4%, 38.9% and 44.4% for 220 kHz, 650 kHz, and 1000 kHz, respectively. An ROI with a thickness deviation of 1.4 mm resulting in the acoustic field developed a sub-focus at 1000 kHz, enlarging the focal area to 2.9 times its size in the free-field, and causing a phase error of 39.0%. In contrast, at 220 kHz and 650 kHz, the field remained largely undistorted, with phase errors of 14.5% and 13.1%, respectively.

The R2 region is located on the superior parietal bone, which is thin (6.4±1.9 mm) compared to the R1, frontal bone region, and R4, posterior parietal bone region (8.0±2.3 mm). However, the RoC (80.3±17.5 mm) is generally larger than in other regions, and larger than the transducer curvature. This caused significant refraction of the sound beam, acted as a diffuser and resulting in a relative lowered peak pressure and broader coverage volume, particular evident at 220 kHz. An ROI with thickness deviation of 1.6 mm, causing phase error of 9.3% at 220 kHz, However, displaying more pronounced phase error of 34.0% and 45.0% at 650 kHz and 1000 kHz respectively.

The R3 region is positioned on the lateral of the parietal bone and temporal bone. Although the structures in this region are on average the thinnest (5.8±1.9 mm), the irregularities of the surfaces and strong curvature (48.5±12.6 mm) contributed to greater phase aberration, reducing the transmission efficiency. One ROI involving with the squamosal suture, with a thickness standard deviation of 2.1 mm, the transmitted pressure was dropped 44% to 56% for three frequencies compared to an adjacent location excluding the squamosal suture with the same mean thickness and a thickness standard deviation is 0.6 mm.

The R4 region, located at the posterior of the parietal bone, had ROIs with a relatively smaller RoC (51.2±7.7 mm) compared to the RoC of the transducers, this often results in which acting as a confocal lens for the sound beam. This configuration reduced acoustic spreading, enabling precise and even tighter focus field control. Results from transmission through this region were overall satisfactory, except for one ROI with a thickness deviation of 1.5 mm, which caused significant distortion in the 1000 kHz field, preventing the formation of a meaningful focus.

### C. Impact of angulation and wave superposition on the transmitted pressure level

FIG. 8 shows the angulation effect on the resultant acoustic field at 650 kHz, the maximum beam depth and width above -6 dB is 37.0 mm and 4.0 mm in free-field- this is the same as when the skull piece was aligned normal to the beam. The depth and width are reduced to 28.0 mm and 3.5 mm at a relative angulation of 15° and 36.5 mm and 3.0 mm at a relative angulation 30°. The propagation direction appeared to be shifted by angulation of the ROI by 5° for a 15° angulation and by 7° for a 30° angulation. The peak intracranial pressure in the acoustic field with the skull piece at 0° and 15° were reduced to -9.8 dB and -10.8 dB of the free-field, respectively, and located close to the geometrical focus. Increasing the incident angle to 30° led to a peak pressure drop of -28.0 dB and shifted the focus 8.0 mm from the geometrical focus.

These preliminary results shown in FIG. 8 demonstrate the importance of keeping the ROI near normal to the transducer aperture, to avoid excessive phase aberration or excitation of shear modes. Additionally, a slight tilt of the transducer appears acceptable, offering a larger acoustic window for LIFU clinical applications, as it can, in some cases, have less impact on the acoustic field than transmitting through skull regions with undesirable morphology.

Although a near normal incidence reduces the phase aberration, this alignment may lead to pronounced wave interactions at the transducer-skull spacing. When employing a coupling medium, typically water or ultrasound gel, which has a significantly lower acoustic impedance to both the skull and transducer, the reflection coefficient at the interface of the media often exceeds 40%[52,54] and 90%,[54] respectively, potentially leading to substantial wave superposition effect. As shown in FIG. 7, concerning a representative skull ROI (thickness of 3.2±0.8 mm, SDR of 0.65±0.25, and RoC of 70.0 mm), significant interference effects are evident at 220 kHz and 650 kHz. The recorded intracranial peak pressure exhibited an oscillation pattern as the transducer-skull spacing increased from 10 to 50 mm. The pressure difference between the waveforms exhibiting constructive and destructive interference is approximately 30% to 40% at 220 kHz and 650 kHz. This effect was mitigated at 1000 kHz, with much lower pressure variation of roughly 10%. The linear fitted results demonstrated a decreasing trend as spacing increased. Specifically, the average pressure reduction for each 10 mm increment was observed to be approximately 2.0%, 3.5%, and 5.0% at 220 kHz, 650 kHz, and 1000 kHz, respectively.

This phenomenon is prominent across the sub-MHz frequency domain and is contingent upon the specific skull morphology. To mitigate standing wave effects, options could include chirped waveforms, random phase-shifting, or using phased-array transducers,[13,55] however, none of these are universal solutions for narrow-bandwidth, single-element transducers that most studies use. Our experiment did not utilise intact skulls, therefore we provide no quantification of standing wave effects within the intracranial domain.

Our preliminary findings strongly indicate that wave superposition effects must be accounted for during FUS delivery to allow for FUS dose at the focal point to be accurately estimated. We have not measured the temperature throughout the experiment, as pressure levels are considerably lower than the safety level proposed by International Transcranial Ultrasonic Stimulation Safety and Standards (ITRUSST).[56]

### D. Future work

When the transducer-skull spacing was maintained between 10 to 20 mm, approximately 0.2 ms for 220 kHz and 650 kHz, and 0.1 ms for 1000 kHz were required for waveform stabilization. For application scenarios using such short time pulses, determining precise insertion loss is challenging, as wave interactions within the skull thickness and between transducer-skull spacing are still stabilizing during this period. The absence of tone burst waveform with a few cycles limited the analysis to system insertion loss, preventing direct comparisons with skull attenuation. Additionally, the close spacing between the skull and transducer complicated distinguishing wave interactions occurring at the transducer-skull spacing from those within the skull thickness.

Future studies could address these limitations by increasing the transducer-skull distance, delaying the superposition of reflected and incident waves, enabling clearer analysis of internal skull wave interactions. Subsequently, adjusting the spacing to clinically relevant ranges would allow a more comprehensive evaluation of the specific effects of transducer-skull spacing on the acoustic transmission.

Although wave superposition effects resulting from changes in transducer-skull spacing are observable across most ROIs, the relationship between the strength of these effects and skull morphology and transducer curvature has not yet been quantitatively studied. In addition, a more comprehensive investigation incorporating scalp phantoms could better replicate realistic transcranial LIFU conditions. Quantifying wave superposition effects for specific transducers would enable future applications to facilitate the precise delivery of acoustic energy tailored to individual skull morphology and transducer placement distance.

Efforts were undertaken to integrate brain phantoms with the Thiel skull for the comprehensive transmission efficiency assessment. Polyvinyl alcohol-based brain phantoms were prepared by following the recommended procedure.[57] The prepared phantom has a sound speed of 1520 m/s and an attenuation coefficient of 0.62±0.11 dB/cm/MHz in the sub-MHz frequency, these parameters are comparable to the reported human brain parameters, which are 1530 m/s and 0.58±0.35 dB/cm/MHz.[1,3,58] However, we were unable to overcome challenges in sealing the skull plus phantom system in the timeframe of this experiment, and this is the subject of ongoing work.

The spatial resolution of the clinical CT imaging system used in this study was sufficient to distinguish between skull layers , whereas insufficient to capture the detailed porous microstructure of the skull. As a result, transmission performance related to the microstructural information of the skull could not be examined, despite being a focus of interest in current studies, such as those focused on transcranial ultrasound imaging.[59]

As previously mentioned, this experiment primarily focused on examining general transcranial LIFU performance. An area of future study would be to assess theoretical target engagement via mapping space within the skull cavity to brain regions, as quantitative experimental treatment of how FUS engages both spatially and temporally with brain targets would be of use to the field.

## V.    CONCLUSION

Comprehensive acoustic field mapping was conducted at frequencies of 220 kHz. 650 kHz and 1000 kHz, and outcomes revealed that the transmitted energy does not correlate to the skull thickness and SDR at 220 kHz, whereas being weakly to moderately correlated at 650 kHz and 1000 kHz. Regarding the size of focal area, a weak correlation with the skull curvature was observed at 220 kHz, with no discernible correlation at 650 kHz and 1000 kHz. Several Thiel results strongly deviated from the historical bound, and we found transmission across ROIs with excessive variations of thickness resulted in substantial field distortion across all tested frequencies. Effects of wave superposition occurring at both transducer-skull spacing and within the skull thickness were noticed during experimentation and investigated, although these preliminary observations have not yet been quantitatively studied. Preliminary angulation tests suggest that an angulation between the skull and transducer below 15° could be tolerable due to insignificant distortion observed. Linear fitted models based on the historical and Thiel skull thickness and historical skull SDR, along with a 95% confidence interval were provided at all three frequencies, notably, the thickness models demonstrated reasonable consistency for both skull types at all frequencies tested, In contrast, the historical skull SDR model exhibited reduced efficacy in predicting Thiel results as the frequency increased. These models could serve as assisting tools for the determination of the output power levels in LIFU treatment planning, therefore, enhancing the precision and efficacy in therapeutic applications.

## ACKNOWLEDGMENTS

Thanks to Tom Gilbertson & Sadaquate Khan at the University of Dundee for helpful discussion, to Alan Webster for significant CT acquisition technical expertise, and to Michelle Cooper facilitating the use of the skull specimens from the Centre for Human Anatomy and Identification (CAHID) at the University of Dundee.

## AUTHOR DECLARATIONS

### Conflict of Interest

The authors have no conflicts of interest to disclose.

**Ethics Approval**

Approval for the use of human tissue specimens in the study was given by the University of Dundee Thiel Advisory Group. This research complies with the Anatomy Act 1984 and the Human Tissue (Scotland) Act 2006.

## DATA AVAILABILITY

The data that support the findings of this study are available on request from the corresponding author. The data are not publicly available due to ethical restrictions.

## REFERENCES

[1] J. Kubanek, "Neuromodulation with transcranial focused ultrasound," Neurosurg Focus **44**(2), (2018).

[2] A. Fomenko, C. Neudorfer, R.F. Dallapiazza, S.K. Kalia, and A.M. Lozano, "Low-intensity ultrasound neuromodulation: An overview of mechanisms and emerging human applications," Brain Stimul **11**(6), 1209–1217 (2018).

[3] W. Legon, L. Ai, P. Bansal, and J.K. Mueller, "Neuromodulation with single-element transcranial focused ultrasound in human thalamus," Hum Brain Mapp **39**(5), 1995–2006 (2018).

[4] S.N. Yaakub, T.A. White, J. Roberts, E. Martin, L. Verhagen, C.J. Stagg, S. Hall, and E.F. Fouragnan, "Transcranial focused ultrasound-mediated neurochemical and functional connectivity changes in deep cortical regions in humans," Nat Commun **14**(1), (2023).

[5] H. Baek, K.J. Pahk, and H. Kim, "A review of low-intensity focused ultrasound for neuromodulation," Biomed Eng Lett **7**(2), 135–142 (2017).

[6] S.T. Brinker, F. Preiswerk, P.J. White, T.Y. Mariano, N.J. McDannold, and E.J. Bubrick, "Focused Ultrasound Platform for Investigating Therapeutic Neuromodulation Across the Human Hippocampus," Ultrasound Med Biol **46**(5), 1270–1274 (2020).

[7] V. Braun, J. Blackmore, R.O. Cleveland, and C.R. Butler, "Transcranial ultrasound stimulation in humans is associated with an auditory confound that can be effectively masked," Brain Stimul **13**(6), 1527–1534 (2020).

[8] C.R. Butler, E. Rhodes, J. Blackmore, X. Cheng, R.L. Peach, M. Veldsman, F. Sheerin, and R.O. Cleveland, "Transcranial ultrasound stimulation to human middle temporal complex improves visual motion detection and modulates electrophysiological responses," Brain Stimul **15**(5), 1236–1245 (2022).

[9] W. Lee, H. Kim, Y. Jung, I.U. Song, Y.A. Chung, and S.S. Yoo, "Image-guided transcranial focused ultrasound stimulates human primary somatosensory cortex," Sci Rep **5**, (2015).

[10] N. Samuel, M.Y.R. Ding, C. Sarica, G. Darmani, I.E. Harmsen, T. Grippe, X. Chen, A. Yang, N. Nasrkhani, K. Zeng, R. Chen, and A.M. Lozano, "Accelerated Transcranial Ultrasound Neuromodulation in Parkinson's Disease: A Pilot Study," Movement Disorders **38**(12), 2209–2216 (2023).

[11] J.L. Sanguinetti, S. Hameroff, E.E. Smith, T. Sato, C.M.W. Daft, W.J. Tyler, and J.J.B. Allen, "Transcranial Focused Ultrasound to the Right Prefrontal Cortex Improves Mood and Alters Functional Connectivity in Humans," Front Hum Neurosci **14**, (2020).

[12] M.W. Sliwinska, S. Vitello, and J.T. Devlin, "Transcranial magnetic stimulation for investigating causal brain-behavioral relationships and their time course," Journal of Visualized Experiments (89), (2014).

[13] S. Yoo, D.R. Mittelstein, R.C. Hurt, J. Lacroix, and M.G. Shapiro, "Focused ultrasound excites cortical neurons via mechanosensitive calcium accumulation and ion channel amplification," Nat Commun **13**(1), (2022).

[14] E. Weinreb, and E. Moses, "Mechanistic insights into ultrasonic neurostimulation of disconnected neurons using single short pulses," Brain Stimul **15**(3), 769–779 (2022).

[15] T. Nandi, B.R. Kop, K. Butts Pauly, C.J. Stagg, and L. Verhagen, "The relationship between parameters and effects in transcranial ultrasonic stimulation," Brain Stimul **17**(6), 1216–1228 (2024).

[16] Y. Tufail, A. Matyushov, N. Baldwin, M.L. Tauchmann, J. Georges, A. Yoshihiro, S.I.H. Tillery, and W.J. Tyler, "Transcranial Pulsed Ultrasound Stimulates Intact Brain Circuits," Neuron **66**(5), 681–694 (2010).

[17] K. Zeng, Z. Li, X. Xia, Z. Wang, G. Darmani, X. Li, and R. Chen, "Effects of different sonication parameters of theta burst transcranial ultrasound stimulation on human motor cortex," Brain Stimul **17**(2), 258–268 (2024).

[18] A. Fomenko, K.H.S. Chen, J.F. Nankoo, J. Saravanamuttu, Y. Wang, M. El-Baba, X. Xia, S.S. Seerala, K. Hynynen, A.M. Lozano, and R. Chen, "Systematic examination of low-intensity ultrasound parameters on human motor cortex excitability and behaviour," Elife **9**, 1–68 (2020).

[19] S. Pichardo, V.W. Sin, and K. Hynynen, "Multi-frequency characterization of the speed of sound and attenuation coefficient for longitudinal transmission of freshly excised human skulls," Phys Med Biol **56**(1), 219–250 (2011).

[20] G. Pinton, J.F. Aubry, E. Bossy, M. Muller, M. Pernot, and M. Tanter, "Attenuation, scattering, and absorption of ultrasound in the skull bone," Med Phys **39**(1), 299–307 (2012).

[21] J. Guo, X. Song, X. Chen, M. Xu, and D. Ming, "Mathematical Model of Ultrasound Attenuation With Skull Thickness for Transcranial-Focused Ultrasound," Front Neurosci **15**, (2022).

[22] T.S. Riis, T.D. Webb, and J. Kubanek, "Acoustic properties across the human skull," Ultrasonics **119**, (2022).

[23] J. Lee, D.-G. Paeng, and K. Ha, "Attenuation of the human skull at broadband frequencies by using a carbon nanotube composite photoacoustic transducer," J Acoust Soc Am **148**(3), 1121–1129 (2020).

[24] A.Y. Ammi, T.D. Mast, I.H. Huang, T.A. Abruzzo, C.C. Coussios, G.J. Shaw, and C.K. Holland, "Characterization of Ultrasound Propagation Through Ex-vivo Human Temporal Bone," Ultrasound Med Biol **34**(10), 1578–1589 (2008).

[25] T. Kirchner, C. Villringer, and J. Laufer, "Evaluation of ultrasound sensors for transcranial photoacoustic sensing and imaging," Photoacoustics **33, 100556**, (2023).

[26] C.W. Connor, and K. Hynynen, "Patterns of thermal deposition in the skull during transcranial focused ultrasound surgery," IEEE Trans Biomed Eng **51**(10), 1693–1706 (2004).

[27] P.J. White, G.T. Clement, and K. Hynynen, "Local frequency dependence in transcranial ultrasound transmission," Phys Med Biol **51**(9), 2293–305 (2006).

[28] F.J. Fry, and J.E. Barger, "Acoustical properties of the human skull," J Acoust Soc Am **63**(5), 1576–90 (1978).

[29] A. Boutet, D. Gwun, R. Gramer, M. Ranjan, G.J.B. Elias, D. Tilden, Y. Huang, S.X. Li, B. Davidson, H. Lu, P. Tyrrell, R.M. Jones, A. Fasano, K. Hynynen, W. Kucharczyk, M.L. Schwartz, and A.M. Lozano, "The relevance of skull density ratio in selecting candidates for transcranial MR-guided focused ultrasound," J Neurosurg **132**(6), 1785–1791 (2020).

[30] M. D'Souza, K.S. Chen, J. Rosenberg, W.J. Elias, H.M. Eisenberg, R. Gwinn, T. Taira, J.W. Chang, N. Lipsman, V. Krishna, K. Igase, K. Yamada, H. Kishima, R. Cosgrove, J. Rumià,

M.G. Kaplitt, H. Hirabayashi, D. Nandi, J.M. Henderson, K.B. Pauly, M. Dayan, C.H. Halpern, and P. Ghanouni, "Impact of skull density ratio on efficacy and safety of magnetic resonance–guided focused ultrasound treatment of essential tremor," J Neurosurg **132**(5), 1392–1397 (2020).

[31] J.-F. Aubry, O. Bates, C. Boehm, K. Butts Pauly, D. Christensen, C. Cueto, P. Gélat, L. Guasch, J. Jaros, Y. Jing, R. Jones, N. Li, P. Marty, H. Montanaro, E. Neufeld, S. Pichardo, G. Pinton, A. Pulkkinen, A. Stanziola, A. Thielscher, B. Treeby, and E. van 't Wout, "Benchmark problems for transcranial ultrasound simulation: Intercomparison of compressional wave models," J Acoust Soc Am **152**(2), (2022).

[32] J. Robertson, E. Martin, B. Cox, and B.E. Treeby, "Sensitivity of simulated transcranial ultrasound fields to acoustic medium property maps," Phys Med Biol **62**(7), 2559–2580 (2017).

[33] H. Niels, "Thirty years of Thiel embalming—A systematic review on its utility in medical research," Clinical Anatomy **35**(7), 987–997 (2022).

[34] P.J. White, S. Palchaudhuri, K. Hynynen, and G.T. Clement, "The effects of desiccation on skull bone sound speed in porcine models," IEEE Trans Ultrason Ferroelectr Freq Control **54**(8), 1708–1710 (2007).

[35] S.J. Estermann, S. Förster-Streffleur, L. Hirtler, J. Streicher, D.H. Pahr, and A. Reisinger, "Comparison of Thiel preserved, fresh human, and animal liver tissue in terms of mechanical properties," Annals of Anatomy **236**, (2021).

[36] Human Tissue (Scotland) Act 2006.

[37] S.N. Yaakub, T.A. White, J. Roberts, E. Martin, L. Verhagen, C.J. Stagg, S. Hall, and E.F. Fouragnan, "Transcranial focused ultrasound-mediated neurochemical and functional connectivity changes in deep cortical regions in humans," Nat Commun **14**(1), (2023).

[38] K. Zeng, Z. Li, X. Xia, Z. Wang, G. Darmani, X. Li, and R. Chen, "Effects of different sonication parameters of theta burst transcranial ultrasound stimulation on human motor cortex," Brain Stimul **17**(2), 258–268 (2024).

[39] J.L. Sanguinetti, S. Hameroff, E.E. Smith, T. Sato, C.M.W. Daft, W.J. Tyler, and J.J.B. Allen, "Transcranial Focused Ultrasound to the Right Prefrontal Cortex Improves Mood and Alters Functional Connectivity in Humans," Front Hum Neurosci **14**, (2020).

[40] W. Legon, L. Ai, P. Bansal, and J.K. Mueller, "Neuromodulation with single-element transcranial focused ultrasound in human thalamus," Hum Brain Mapp **39**(5), 1995–2006 (2018).

[41] K.W.K. Tsai, J.C. Chen, H.C. Lai, W.C. Chang, T. Taira, J.W. Chang, and C.Y. Wei, "The Distribution of Skull Score and Skull Density Ratio in Tremor Patients for MR-Guided Focused Ultrasound Thalamotomy," Front Neurosci **15**, (2021).

[42] A. Boutet, D. Gwun, R. Gramer, M. Ranjan, G.J.B. Elias, D. Tilden, Y. Huang, S.X. Li, B. Davidson, H. Lu, P. Tyrrell, R.M. Jones, A. Fasano, K. Hynynen, W. Kucharczyk, M.L. Schwartz, and A.M. Lozano, "The relevance of skull density ratio in selecting candidates for transcranial MR-guided focused ultrasound," J Neurosurg **132**(6), 1785–1791 (2020).

[43] M. D'Souza, K.S. Chen, J. Rosenberg, W.J. Elias, H.M. Eisenberg, R. Gwinn, T. Taira, J.W. Chang, N. Lipsman, V. Krishna, K. Igase, K. Yamada, H. Kishima, R. Cosgrove, J. Rumià, M.G. Kaplitt, H. Hirabayashi, D. Nandi, J.M. Henderson, K.B. Pauly, M. Dayan, C.H. Halpern, and P. Ghanouni, "Impact of skull density ratio on efficacy and safety of magnetic resonance–guided focused ultrasound treatment of essential tremor," J Neurosurg **132**(5), 1392–1397 (2020).

[44] ISO/IEC, "IEC 62127-1 Ultrasonics - Hydrophones - Part 1: Measurement and characterization of medical ultrasonic fields," (2013).

[45] ISO/IEC, "IEC 62127-2 Ultrasonics - Hydrophones - Part 2: Calibration for ultrasonic fields up to 40 MHz," ISO/IEC, (2007).

[46] ISO/IEC, "IEC 61828 Ultrasonics - Transducers - Definitions and measurement methods regarding focusing for the transmitted fields," ISO/IEC, (2020).

[47] ISO/IEC, "IEC 61689 Ultrasonics - Physiotherapy systems - Field specifications and methods of measurement in the frequency range 0,5 MHz to 5 MHz," ISO/IEC, (2022).

[48] ISO/IEC, Ultrasonics – Power Measurement –Radiation Force Balances and Performance Requirements (2006).

[49] O.A. Sapozhnikov, S.A. Tsysar, V.A. Khokhlova, and W. Kreider, "Acoustic holography as a metrological tool for characterizing medical ultrasound sources and fields," J Acoust Soc Am **138**(3), 1515–1532 (2015).

[50] M. Chen, C. Peng, H. Wu, C.C. Huang, T. Kim, Z. Traylor, M. Muller, P.Y. Chhatbar, C.S. Nam, W. Feng, and X. Jiang, "Numerical and experimental evaluation of low-intensity transcranial focused ultrasound wave propagation using human skulls for brain neuromodulation," Med Phys **50**(1), 38–49 (2023).

[51] G. Pinton, J.F. Aubry, E. Bossy, M. Muller, M. Pernot, and M. Tanter, "Attenuation, scattering, and absorption of ultrasound in the skull bone," Med Phys **39**(1), 299–307 (2012).

[52] D.N. White, G.R. Curry, and R.J. Stevenson, *THE ACOUSTIC CHARACTERISTICS OF THE SKULL\** (Pergamon Press, 1978).

[53] D. Attali, T. Tiennot, M. Schafer, E. Fouragnan, J. Sallet, C.F. Caskey, R. Chen, G. Darmani, E.J. Bubrick, C. Butler, C.J. Stagg, M. Klein-Flügge, L. Verhagen, S.S. Yoo, K.B. Pauly, and J.F. Aubry, "Three-layer model with absorption for conservative estimation of the maximum acoustic transmission coefficient through the human skull for transcranial ultrasound stimulation," Brain Stimul **16**(1), 48–55 (2023).

[54] V.T. Rathod, "A review of acoustic impedance matching techniques for piezoelectric sensors and transducers," Sensors (Switzerland) **20**(14), 1–65 (2020).

[55] S.C. Tang, and G.T. Clement, "Standing-wave suppression for transcranial ultrasound by random modulation," IEEE Trans Biomed Eng **57**(1), 203–205 (2010).

[56] J.-F. Aubry, D. Attali, M. Schafer, E. Fouragnan, C. Caskey, R. Chen, G. Darmani, E.J. Bubrick, J. Sallet, C. Butler, C. Stagg, M. Klein-Flügge, S.-S. Yoo, B. Treeby, L. Verhagen, and K. Butts Pauly, ITRUSST Consensus on Biophysical Safety for Transcranial Ultrasonic Stimulation (n.d.).

[57] S. Taghizadeh, C. Labuda, and J. Mobley, "Development of a Tissue-Mimicking Phantom of the Brain for Ultrasonic Studies," Ultrasound Med Biol **44**(12), 2813–2820 (2018).

[58] F. Duck, Physical Properties of Tissues: A Comprehensive Reference Book (Academic press, 2013).

[59] B. Jing, and B.D. Lindsey, "Effect of Skull Porous Trabecular Structure on Transcranial Ultrasound Imaging in the Presence of Elastic Wave Mode Conversion at Varying Incidence Angle," Ultrasound Med Biol **47**(9), 2734–2748 (2021).